\documentclass[3p,times]{elsarticle}

\RequirePackage{lineno}
\usepackage{ecrc}

\usepackage{amsmath,amsthm,amssymb,commath,bm,mathrsfs}
\usepackage[normalem]{ulem}
\usepackage{listings,color,graphicx}
\usepackage{float}
\usepackage{subfloat}
\usepackage{subfig}
\usepackage{xspace}
\usepackage{fixltx2e}
\usepackage{booktabs}
\usepackage{hyperref}
\usepackage{algpseudocode}
\usepackage{algorithm}
\usepackage{placeins}
\usepackage[utf8]{inputenc}
\usepackage[T1]{fontenc}

\newcommand{\FPROG}{\mathcal{F}}
\newcommand{\CPROG}{\mathcal{G}}
\newcommand{\Tau}{\mathrm{T}}
\newcommand{\AlgAbb}{\texttt{SPASD}} 

\volume{}
\firstpage{1}
\journalname{arXiv.org}
\runauth{A. Blumers et al.}

\begin{document}

\begin{frontmatter}
\title{Supervised parallel-in-time algorithm for long-time Lagrangian simulations of stochastic dynamics: Application to hydrodynamics}
\author[a]{Ansel L. Blumers}
\ead{ansel\_blumers@brown.edu}
\author[b]{Zhen Li\corref{cor1}}
\ead{zhen\_li@brown.edu}
\author[b]{George Em Karniadakis\corref{cor1}}
\ead{george\_karniadakis@brown.edu}
\cortext[cor1]{Corresponding authors}
\address[a]{Department of Physics, Brown University, Providence, RI 02912, USA}
\address[b]{Division of Applied Mathematics, Brown University, Providence, RI 02912, USA}

\begin{abstract}
Lagrangian particle methods based on detailed atomic and molecular models are powerful computational tools for studying the dynamics of microscale and nanoscale systems. However, the maximum time step is limited by the smallest oscillation period of the fastest atomic motion, rendering long-time simulations very expensive.
To resolve this bottleneck, we propose a supervised {\em parallel-in-time} algorithm for stochastic dynamics (\AlgAbb) to accelerate long-time Lagrangian particle simulations. Our method is inspired by bottom-up coarse-graining projections that yield mean-field hydrodynamic behavior in the continuum limit. Here as an example, we use the dissipative particle dynamics (DPD) as the Lagrangian particle simulator that is supervised by its macroscopic counterpart, i.e., the Navier-Stokes simulator. The low-dimensional macroscopic system (here, the Navier-Stokes solver) serves as a predictor to supervise the high-dimensional Lagrangian simulator, in a predictor-corrector type algorithm. The results of the Lagrangian simulation then correct the mean-field prediction and provide the proper microscopic details (e.g., consistent fluctuations, correlations, etc.).
The unique feature that sets \AlgAbb~apart from other multiscale methods is the use of a low-fidelity macroscopic model as a predictor. The macro-model can be approximate and even inconsistent with the microscale description, but \AlgAbb~anticipates the deviation and corrects it internally to recover the true dynamics.
We first present the algorithm and analyze its theoretical speedup, and subsequently we present the accuracy and convergence of the algorithm for the time-dependent plane Poiseuille flow, demonstrating that \AlgAbb~converges exponentially fast over iterations, irrespective of the accuracy of the predictor. Moreover, the fluctuating characteristics of the stochastic dynamics are identical to the unsupervised (serial in time) DPD simulation. We also compare the performance of \AlgAbb~to the conventional spatial decomposition method, which is one of the most parallel-efficient methods for particle simulations. We find that the parallel efficiency of \AlgAbb~and the conventional spatial decomposition method are similar for a small number of computing cores, but for a large number of cores the performance of \AlgAbb~is superior. Furthermore, \AlgAbb~can be used in conjunction with spatial decomposition for enhanced performance. Lastly, we simulate a two-dimensional cavity flow that requires more iterations to converge compared to the Poiseuille flow, and we observe that \AlgAbb~converges to the correct solution. Although a DPD solver is used to demonstrate the results, \AlgAbb~is a general framework and can be readily applied to other Lagrangian approaches including molecular dynamics and Langevin dynamics.
\end{abstract}


\end{frontmatter}

\hyphenpenalty=800
\tolerance=100

\section{Introduction}
Inherent stochastic fluctuations are important in many cellular and subcellular processes in biological systems, as well as in physical systems (soft matter, plasma dynamics, etc.) with timescales across many orders of magnitude. For such systems, a macroscopic deterministic description built on the continuum hypothesis is invalid~\cite{2007Erban}, and therefore Lagrangian methods based on detailed atomic and molecular models are required~\cite{2012Perez}. However, because the atomistic details are explicitly modeled, the maximum time step is limited by the smallest oscillation period of the fastest atomic motion, which are typically of the order of several femtoseconds~\cite{1999Feenstra}. Consequently, the physical time scales accessible to those simulations are too short to address interesting long-time dynamics. This huge disparity in temporal scales, which can also happen at larger values of the smallest timescales, has not been addressed adequately in the computational literature on multiscale physical and biological systems.
Consider, for example, the problem of thrombus formation caused by platelet margination and adhesion to the arterial wall, where the characteristic time for platelet activation is a few microseconds. Lagrangian platelet models can only simulate the platelet dynamics up to a few minutes~\cite{2009Pivkin,2017Yazdani}, which is far below the time scale for those {\em in vivo} processes that usually require at least several hours or even days~\cite{2010Killer,2012Schriefl}. With a time step of the order of $10^{-6}~s$, a clinically meaningful result for at least one-day time integration may require about $10^{11}$ time steps with explicit platelet models~\cite{2008Filipovic}, which is computationally prohibitive.
Another example is the simulation of protein folding dynamics~\cite{freddolino2010challenges}. In an {\em ab initio} molecular dynamics~(MD) simulation, the protein molecule and surrounding solvents are treated as classical particles interacting through empirical energy functions. The MD trajectories provide extremely high spatial and temporal resolutions and can identify key intermediates of the folding processes. However, its computational cost hampers the study of large proteins due to their large number of degrees of freedom, greatly increasing the simulation time required to observe a single folding event, which is on the order of tens of microseconds to milliseconds~\cite{prigozhin2013microsecond}. For timescale on the order of milliseconds, MD simulations are challenging even for the specially designed ASICS machine - the longest continuous-trajectory atomic-scale simulation is a $2.936$ millisecond simulation of NTL9 at $355~K$~\cite{Lindorff-Larsen517}. Major conformational transitions triggered by enzymatic reactions occur on the millisecond to second time scale or longer~\cite{elber2016perspective}. Without algorithmic improvements, long-time integration of protein folding via {\em ab initio} MD simulations is a formidable problem.

The complexity due to the huge number of spatio-temporal degrees of freedom needed to resolve atomic or molecular details requires large amounts of computational resources and simulation time to capture the underlying kinetics and dynamics. A typical strategy to leverage parallel computing is dividing the spatial domain into many smaller subdomains so that all subdomains can be processed concurrently by many computing cores. We will refer to this as ``conventional spatial decomposition'' approach hereafter. Although it is relatively easy-to-implement and very effective, the strong scaling eventually plateaus as the number of unknowns per core becomes too small~\cite{blumers2017gpu}. The lack of an efficient and stable numerical method remains a serious bottleneck for long-time simulations. Time parallel integration methods can break the bottleneck by decomposing the time domain and solving these subdomains parallel-in-time. Since the inception of time parallel integration methods over 50 years ago~\cite{nievergelt1964parallel}, many have improved and proposed new parallel-in-time methods. Parareal, proposed by Lions et al.~\cite{lions2001resolution}, is a popular parallel-in-time scheme and has demonstrated its versatility in various applications. It follows a predictor-corrector paradigm in which the prediction is carried out with an inexpensive solver while the correction is performed with an expensive but parallelizable solver. The prediction is then corrected over iterations, resulting in a refined solution, which then acts as an initial condition in time for the next iteration. The traditional Parareal scheme has been applied to a wide range of problems, including fluid-structure interactions~\cite{FarhatEtAl2003}, reservoir simulation~\cite{garrido2005convergent}, approximation to Navier-Stokes equations~\cite{fischer2005parareal}, turbulent plasma flow~\cite{samaddar2010parallelization}, and even to fractional partial differential equations~\cite{2015Xu}. While most applications employ continuum solvers, a few utilize other methods such as particle solvers~\cite{baffico2002parallel,bylaska2013extending,astorino2012multiscale,speck2012massively,frantziskonis2009time,legoll2013micro}. Nevertheless, in most published works the fine and coarse propagators solve the same governing equation in an umbrella model, but with different spatial-temporal discretizations. There are a few exceptions in which the coarse propagator operates on a simplified mathematical model~\cite{blouza2010parallel,engblom2009parallel,he2010reduced,astorino2012multiscale}. However, the acceleration of Lagrangian stochastic particle solvers employing parallel-in-time strategy has yet to be developed.

In the present work, we propose a {\bf S}upervised {\bf P}arallel-in-time {\bf A}lgorithm for {\bf S}tochastic {\bf D}ynamics (\AlgAbb), which aims to significantly accelerate stochastic Lagrangian solvers for long-time simulations. Based on the bottom-up coarse-graining philosophy, stochastic particle models such as dissipative particle dynamics (DPD) converge to continuum macroscopic models in the scale limit~\cite{1995Espanol,2001Zwanzig}. The macroscopic system can then serve as a predictor to supervise the high-dimensional stochastic Lagrangian simulation. Even though the governing equations of the macroscopic model, generally in the form of partial differential equations, are different from those of the microscopic model, the macroscopic model can capture the correct mean-field behavior of the microscopic system in the continuum limit. In particular, an inexpensive continuum solver, also known as the coarse propagator, solves the macroscopic model in serial, and an expensive but parallelizable solver, also known as the fine propagator, resolves the molecular details by performing stochastic microscopic simulations. In the examples, we will show how the coarse propagator can provide a rough prediction of the mean-field hydrodynamics that supervises the fine propagator in the time domain, as well as how the fine propagator corrects the prediction iteratively. For demonstration, we employ DPD to model the stochastic microscopic dynamics and the Navier-Stokes equations to obtain mean-field behavior in the continuum~\cite{1995Espanol,2001Zwanzig}. It is worth noting that \AlgAbb~is a scalable and stable parallel-in-time algorithm applicable to many popular stochastic Lagrangian solvers, e.g., MD, DPD, smoothed DPD and Langevin dynamics. Most importantly, the strong scaling of \AlgAbb~does not plateau like conventional spatial decomposition methods, and \AlgAbb~can be implemented in conjunction with spatial decomposition methods to achieve further speedup.

Considering \AlgAbb~as a multiscale method, one could argue that \AlgAbb~resembles methods such as multigrid, heterogeneous multiscale method, and equation-free approach. However, there are clear distinctions. Categorically, \AlgAbb~belongs to the class of concurrent multiscale modeling, for which the macro- and micro-model are solved concurrently. Structurally, \AlgAbb~echoes the extended multigrid methods (EMGM) \cite{brandt2002multiscale}, in which the effective problems are solved with different types of models corresponding to different levels of physical representation. To bridge the scales, EMGM employs techniques that are similar to \AlgAbb~- interpolation, relaxation and restriction, and relies on iterative refinement for convergence. However, in contrast with \AlgAbb~, EMGM is strictly serial in time. There have been some attempts for developing parallel-in-time multigrid methods~\cite{horton1992time}, but the extension to EMGM has yet to be properly explored to the best of our knowledge.

The heterogeneous multiscale method (HMM) views the connection between the micro- and macro-model from a different perspective. In HMM, one has to first guess the form of the macro-model, which can be incomplete~\cite{weinan2003heterogeneous}, e.g. missing the true constitutive law in complex materials. The key is to extract the required information from the micro-model to solve the incomplete macro-model. As expected, the macro-state provides the constraints for setting up the micro-state. In this regard, the difference between HMM and \AlgAbb~is easy to spot: the macro-model is complete and serves as a predictor in \AlgAbb~. In particular, guessing the wrong values of the macro-model parameters in \AlgAbb~is acceptable because the ramification is corrected by the algorithm, whereas guessing the wrong form of the macro-model is likely going to lead to wrong results in HMM~\cite{weinan2003heterogeneous}.

Another well-known multiscale methodology is the equation-free approach (EFA). EFA is intended to be used when a macroscopic description is unavailable in closed form~\cite{kevrekidis2003equation}. In EFA, microscopic simulations are performed only in small spatial domains over a short time period, commonly known as patches. The evolution of the macroscopic fields is then macroscopically interpolated across the patches. In contrast, microscopic simulations in \AlgAbb~are performed concurrently on the entire temporal domain. Furthermore, we use a macro-model to best describe the complex microscopic dynamics at a coarse level. Due to the reduced representation of the true dynamics, the macro-model has a closed-form, and it is solvable with a fast solver. 

In conclusion, treating the macro-model merely as a predictor is unique to \AlgAbb~. That is to say, the macro-model can be inaccurate and inconsistent with the microscale description. Because the parallel-in-time algorithm anticipates the deviation from the true (microscopic) dynamics, \AlgAbb~can recover the true dynamics, as demonstrated in this work.

The remainder of this paper is organized as follows. In section~\ref{sec:2} we describe \AlgAbb~and provide an analysis of its theoretical speedup. In Section~\ref{sec:3} we demonstrate the implementation of \AlgAbb~for a stochastic Lagrangian simulation supervised by its mean-field approximation. Subsequently, we present quantitative results including accuracy, convergence rate and parallel efficiency for an one-dimensional time-dependent problem. Furthermore, we apply \AlgAbb~in a two-dimensional cavity flow and analyzed the results. Lastly, the paper ends with a brief summary and discussion in Section~\ref{sec:SumDis}.

\section{Algorithm}\label{sec:2}
Given that \AlgAbb~is inspired by the traditional Parareal algorithm~\cite{lions2001resolution}, we will first briefly summarize the traditional Parareal and then introduce the proposed algorithm, \AlgAbb. The speedup offered by \AlgAbb~is determined by the complexity of projection, mapping and filtering operations as described below. In order to retain generality, we present a theoretical speedup in which the walltimes of those operations are represented symbolically. The theoretical speedup is then compared to the speedup  of conventional spatial decomposition method in the context of distributed computing.

\subsection{Traditional Parareal Algorithm}
For the sake of clarity, let us assume that we have reduced the original problem to the following ordinary differential equation (ODE): ${du}/{dt}=f(u), \; t~\in~[0,T_\text{final}]$; see Table \ref{tab:1} for the notation. Let $U^n_k$ be an approximation to the exact solution $u(n\Delta T)$ at time $n\Delta T$ in $k^\text{th}$ iteration, where $\Delta T$ denotes the length of a time subdomain. $N=T_\text{final}/\Delta T$ is the number of time subdomains.
Let $\FPROG$ be the expensive fine propagator that accurately approximates $u(t)$, and $\CPROG$ be a less accurate but inexpensive coarse propagator. $\Delta t_f$ and $\Delta t_c$ represent the time step of the fine and coarse propagators, respectively. We note that, although $\Delta t_c$ is usually equal to the length of a time subdomain $\Delta T$, it can be less than $\Delta T$ as demonstrated in later sections, i.e., $\Delta t_c \le \Delta T$.
Summarized in Algorithm~\ref{alg:OrigParareal}, the traditional Parareal algorithm~\cite{lions2001resolution} places few constraints on the fine and coarse propagators. Typically, the coarse propagator employs the same numerical scheme as the fine solver, but with a coarser spatio-temporal discretization. Alternatively, they can be different numerical schemes with distinct convergence properties~\cite{engblom2009parallel,mitran2010time}. Nevertheless, both the fine and coarse propagators operate on the same underlying governing equations in most applications.

\begin{algorithm}[h!]
\caption{Traditional Parareal algorithm}
\begin{algorithmic}[1]
\State Initialization:

Compute initial iteration with Coarse propagator: $U^{n+1}_0 = \CPROG(U^n_0)$ for all $0\le n \le N$.

\State Assume sequence $\{U^n_k\}$ is known for some $0\leq n\leq N$ and $k\geq0$:

Correction:

\hspace{\algorithmicindent} (a) Advance with Coarse propagator in serial:  $\CPROG(U^n_k)$ for all $0\le n \le N-1$.

\hspace{\algorithmicindent} (b) Advance with Fine propagator in parallel:  $\FPROG(U^n_k)$ for all $0\le n \le N-1$.

\hspace{\algorithmicindent} (c) Compute correction:  $\delta^n_k = \FPROG(U^n_k) - \CPROG(U^n_k)$ for all $0\le n \le N-1$.

Prediction:

\hspace{\algorithmicindent} Advance with Coarse propagator in serial:  $\CPROG(U^n_{k+1})$ for all $0\le n \le N-1$.

Refinement:

\hspace{\algorithmicindent} Combine the correction and prediction terms: $U^{n+1}_{k+1} = \CPROG(U^n_{k+1}) + \FPROG(U^n_k) - \CPROG(U^n_k) $

\State Repeat \textit{Step 2} to compute $U^{n}_{k+2}$ for all $1\le n \le N$ until a termination-condition is satisfied.

\end{algorithmic}
\label{alg:OrigParareal}
\end{algorithm}

\begin{table}[h!]
\begin{center}
  \captionof{table}{Notation for selected variables.} \label{tab:1}
  \begin{tabular}{ l  l }
    \hline
    $\Delta T$   & Length of a time subdomain \\
    $\Delta t_c$ & Time step of a coarse propagator  \\
    $\Delta t_f$ & Time step of a fine propagator \\
    $K$ & Number of iterations \\
    $N$ & Number of time subdomains \\
    $\FPROG$ & Fine propagator \\
    $\CPROG$ & Coarse propagator \\
    $\mathscr{F}$ & Filtering operator \\
    $\mathscr{R}$ & Mapping operator \\
    $\mathscr{P}$ & Projection operator \\
    $\Tau^\text{serial}_\text{SD}$ & Total walltime of a serial simulation using conventional SD \\
    $\Tau^\text{\AlgAbb}$ & Total walltime of a simulation using \AlgAbb \\
    $\Tau^\text{f}$ & Walltime taken by fine propagations per iteration \\
    $\Tau^\text{c}$ & Walltime taken by coarse propagations per iteration \\
    $\Tau^\mathscr{F}$ & Walltime taken by filtering operations per iteration \\
    $\Tau^\mathscr{R}$ & Walltime taken by mapping operations per iteration \\
    $\Tau^\mathscr{P}$ & Walltime taken by projection operations per iteration \\
    $\tau^\text{f}$ & Walltime of one fine propagation \\
    $\tau^\text{c}$ & Walltime of one coarse propagation \\
    $\tau^\mathscr{F}$ & Walltime of one filtering operation \\
    $\tau^\mathscr{R}$ & Walltime of one mapping operation \\
    $\tau^\mathscr{P}$ & Walltime of one projection operation \\
    \hline
  \end{tabular}
\end{center}
\end{table}

\subsection{Supervised Parallel-in-time Algorithm for Stochastic Dynamics (\AlgAbb)~\label{sec:pasd}}
\begin{figure}[b!]
	\centering
	\includegraphics[width=0.54\textwidth]{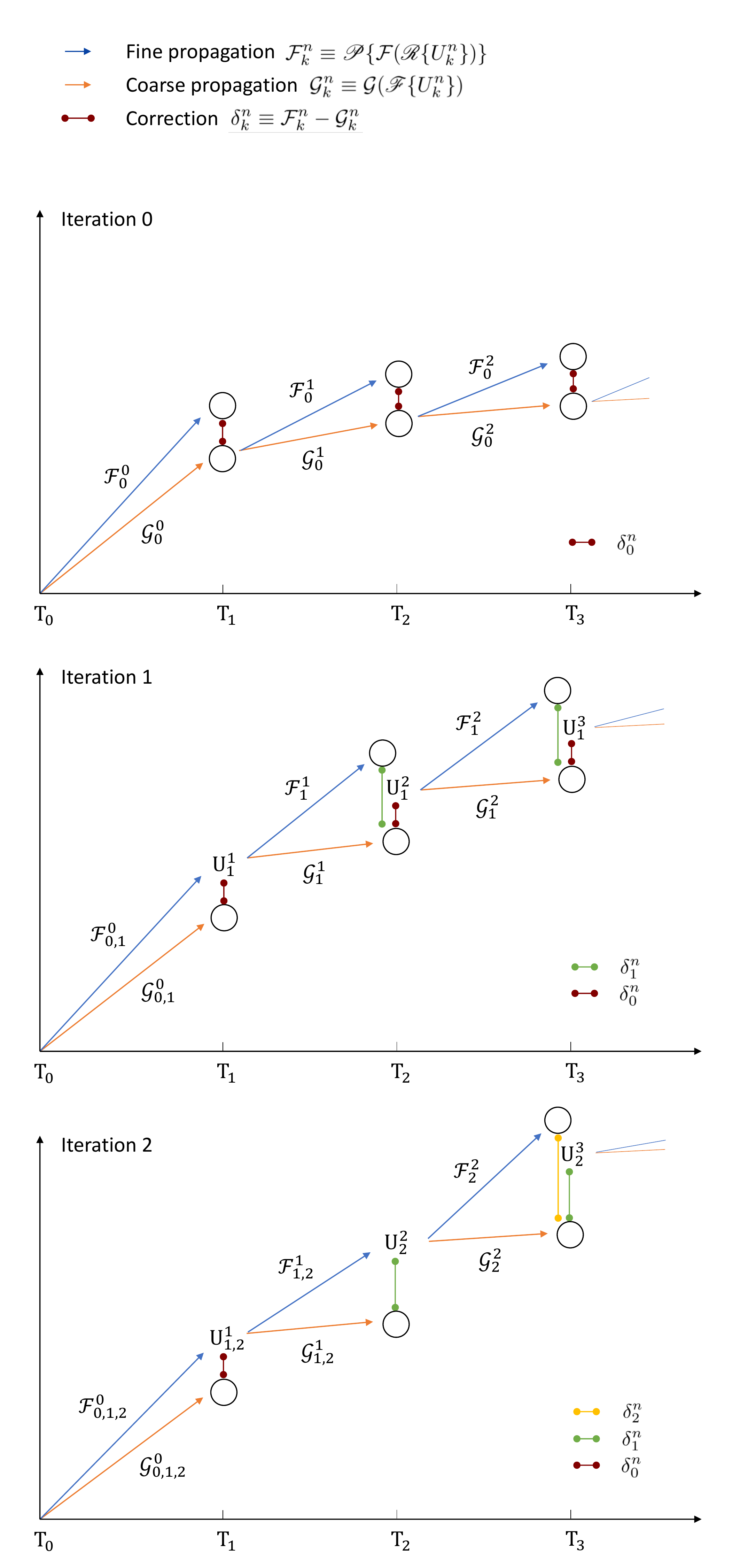}
    \caption{A graphical representation of our proposed algorithm \AlgAbb. Unlike fine $\FPROG$ and coarse $\CPROG$ propagations in the traditional Parareal algorithm, $\FPROG$ and $\CPROG$ in \AlgAbb~encompass other operations, i.e., mapping, projection, filtering. The mapping and projection operators in $\FPROG \equiv \mathscr{P}\{\FPROG(\mathscr{R}\{U^n_k\})\}$ serve to link the two scales in the multiscale context, while the filtering operator in $\CPROG \equiv \CPROG(\mathscr{F}\{U^n_k\})$ is designed to suppress the stochastic noise in the model. In Iteration 1, $U^1_1$, $U^2_1$ and $U^3_1$ are computed from their respective $\mathcal{F}^n_k$ and $\mathcal{G}^n_k$. Because the initial state is fixed, $\mathcal{F}^0_0$ and $\mathcal{F}^0_1$ are identical. Consequentially, $U^1_1$ is equivalent to $U^1_2$ in Iteration 2.}
	\label{fig:MultiParareal}
\end{figure}

In order to accelerate particle simulations, we use a mean-field approximation as the macroscopic model whose underlying governing equations are different from those of the microscopic model. Summarized in Algorithm~\ref{alg:MultiParareal} and represented graphically in Fig.~\ref{fig:MultiParareal}, propagators in \AlgAbb~solve their respective sets of governing equations reflecting the model. As a consequence, a solution representing the macroscopic state is different from the one representing the microscopic state. A macroscopic solution from a coarse propagation can be mapped to a microscopic state via a mapping operation (denoted by the mapping operator $\mathscr{R}$). Conversely, a macroscopic solution can be obtained from a microscopic state via a projection operation (denoted by the projection operator $\mathscr{P}$). The projection and mapping processes are the core challenges for virtually all types of multiscale algorithms. There is no universal method bridging the scales, and the realization of those processes should be problem-dependent. During our numerical experiments, we tested some commonly implemented methods in multiscale such as weighted kernel sampling, linearly interpolation, and nearest neighbor interpolation. For our purpose, we found that the choice of method does not impact the refined results, which is partially due to the fact that the information lost in those processes can be regained by the fine propagator.

Noise filtering, denoted by operator $\mathscr{F}$, is crucial to the success of \AlgAbb~because in stochastic simulations the noise would accumulate with time if left unattended. The accumulation of noise will cause the refined solution to diverge from the true solution, as demonstrated in Section~\ref{sec:ImpactOfFiltering}.

\begin{algorithm}[h!]
\caption{\AlgAbb}
\begin{algorithmic}[1]
\State Initialization:

Compute initial iteration with Coarse propagator: $U^{n+1}_0 = \CPROG(U^0_0)$ for all $0\le n \le N$.

\State Assume sequence $\{U^n_k\}$ is known for some $0\leq n\leq N$ and $k\geq0$:

Correction:

\hspace{\algorithmicindent} (a) Filter solution to remove noise: $\mathscr{F}\{U^n_k\}$

\hspace{\algorithmicindent} (b) Advance with Coarse propagator in serial:  $\CPROG(\mathscr{F}\{U^n_k\})$ for all $0\le n \le N-1$.

\hspace{\algorithmicindent} (c) Map the macroscopic state to microscopic space:  $\mathscr{R}\{U^n_k\}$

\hspace{\algorithmicindent} (d) Advance with Fine propagator in parallel:  $\FPROG(\mathscr{R}\{U^n_k\})$ for all $0\le n \le N-1$.

\hspace{\algorithmicindent} (e) Project the microscopic state to macroscopic state:  $\mathscr{P}\{\FPROG(\mathscr{R}\{U^n_k\})\}$

\hspace{\algorithmicindent} (f) Compute the correction: $\delta^n_k \equiv \mathscr{P}\{\FPROG(\mathscr{R}\{U^n_k\})\} - \CPROG(\mathscr{F}\{U^n_k\})$ for all $0\le n \le N-1$.

Prediction:

\hspace{\algorithmicindent} Advance with Coarse propagator in serial:  $\CPROG(\mathscr{F}\{U^n_{k+1}\})$ for all $0\le n \le N-1$

Refinement:

\hspace{\algorithmicindent} Combine the correction and the prediction terms:

\hspace{\algorithmicindent} $U^{n+1}_{k+1} = \CPROG(\mathscr{F}\{U^n_{k+1}\}) + \mathscr{P}\{\FPROG(\mathscr{R}\{U^n_k\})\} - \CPROG(\mathscr{F}\{U^n_k\})$  for all $0\le n \le N-1$.

\State Repeat \textit{Step 2} to compute $U^{n}_{k+2}$ for all $1\le n \le N$ until a termination-condition is satisfied.

\end{algorithmic}
\label{alg:MultiParareal}
\end{algorithm}

As with all iterative schemes, the refinement process stops when a termination-condition is satisfied. Because the magnitude of the stochastic noise varies from application to application, the idea of a universal condition is imprudent. We thus propose a practical condition that can be tuned based on the application:
\begin{equation}
\frac{ \Big\lVert \CPROG(\mathscr{F}\{U^n_{k+1}\}) - \CPROG(\mathscr{F}\{U^n_k\}) \Big\rVert_{l_1} } {\Big\lVert \CPROG(\mathscr{F}\{U^n_{k+1}\}) \Big\rVert_{l_1} } \equiv C_{\text{TC}} < C_{\text{Tolerance}} ,
\label{eqn:term_condition}
\end{equation}
where $C_{\text{Tolerance}}$ is a user-defined tolerance, which should be determined by considering the magnitude of the intrinsic stochastic noise in the system.

\subsection{Theoretical Speedup} \label{subsec:TheoSpeedup}
\begin{figure}[t!]
	\centering
    \addtolength\tabcolsep{14pt}
    \begin{tabular}{c c}
	\subfloat[]{\includegraphics[width=0.45\textwidth]{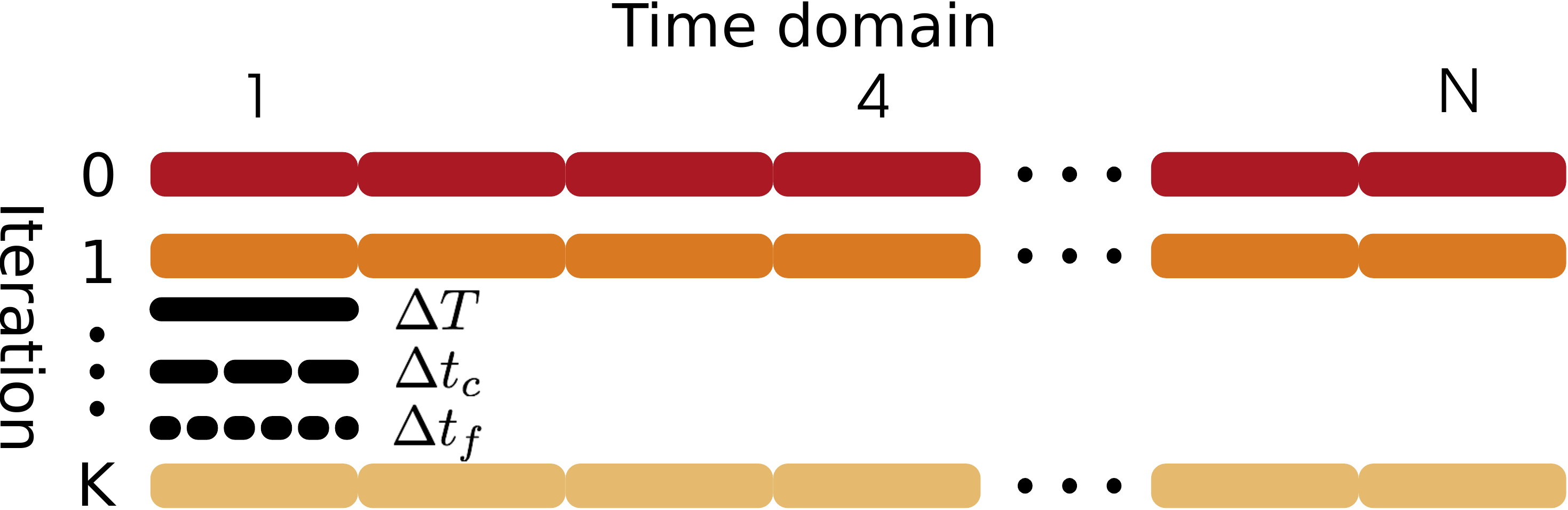}\label{domain}} &
    \subfloat[]{\includegraphics[width=0.4\textwidth]{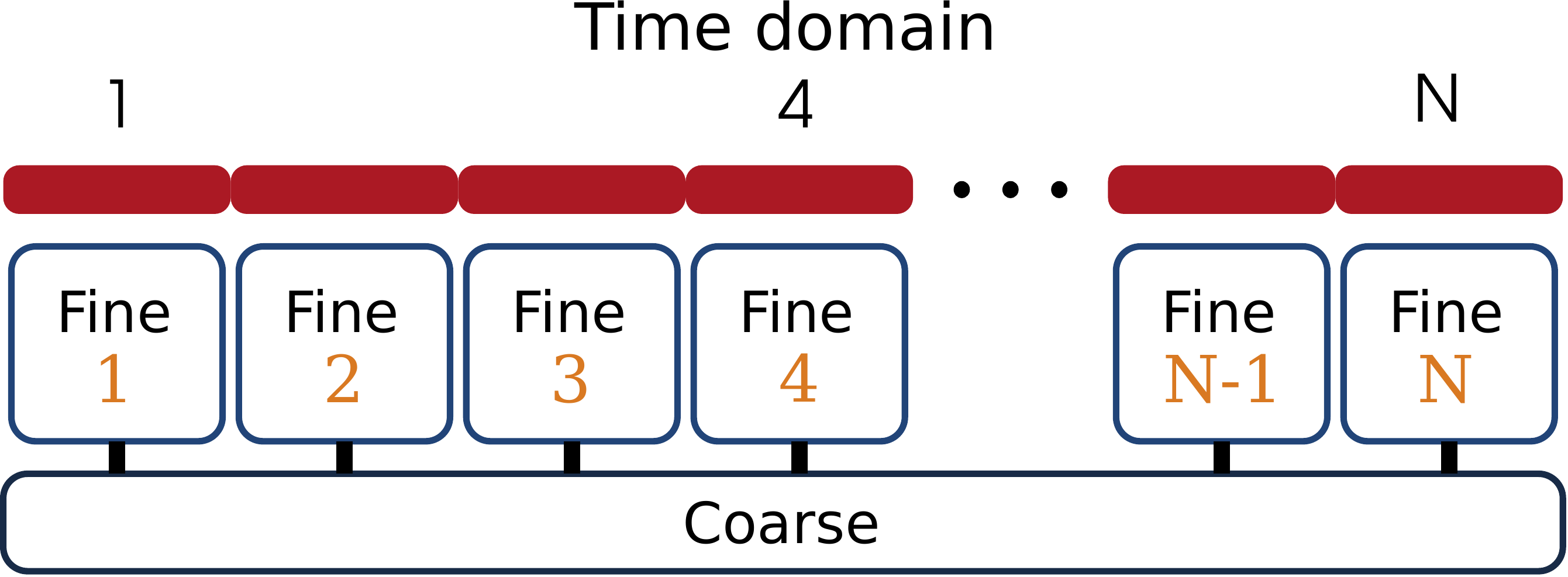}\label{domainfc}}
	\end{tabular}
    \caption{(a) A schematic representation of the temporal decomposition in \AlgAbb. The entire time domain is divided into $N$ subdomains, each of which is represented by a segment. As described in Algorithm~\ref{alg:MultiParareal}, the first iteration is the initialization stage, which involves only the coarse propagation. (b) Each fine propagator performs computation on each subdomain, and only one coarse propagator is used for the entire time domain. Let $\tau^c$ and $\tau^f$ denote the walltime taken by each coarse and fine propagation, respectively. For the sequential coarse propagation, the total walltime spent on one iteration is $N \cdot \tau^c$. Since the fine propagators can be executed concurrently, the walltime taken by fine propagation is simply $\tau^f$ for one iteration. In summary, the total walltime spent on each iteration is thus $N \cdot \tau^c + \tau^f$ and $K \cdot (N \cdot \tau^c + \tau^f)$ for $K$ iterations.}
	\label{fig:time_decomp}
\end{figure}

The total walltime can be expressed as $\Tau^\text{\AlgAbb} = K \cdot ( \Tau^\text{f} + \Tau^\text{c} + \Tau^\mathscr{F} + \Tau^\mathscr{R} + \Tau^\mathscr{P} )$, where $K$ is the number of iterations to convergence. The walltimes taken by the coarse and fine propagators per iteration are denoted by $\Tau^c$ and $\Tau^f$, respectively. The rest of the terms $\Tau^\mathscr{F}$, $\Tau^\mathscr{R}$ and $\Tau^\mathscr{P}$ denote walltime spent on noise-filtering, up-scaling and down-scaling, respectively. In accordance with the notation in Fig.~\ref{fig:time_decomp}, let $N$ be the number of concurrent fine propagations. $\Tau^\text{c}$ and $\Tau^\text{f}$ can therefore be expressed as
\begin{gather}
\Tau^\text{c} = N \cdot \tau^\text{c}; \\
\Tau^\text{f} = \tau^\text{f},
\end{gather}
where $\tau^\text{f}$ is the walltime taken by the fine solver for one time-subdomain, and $\tau^\text{c}$ is the walltime taken by the coarse solver for the same time-subdomain. $\Tau^\mathscr{F}$, $\Tau^\mathscr{R}$ and $\Tau^\mathscr{P}$ can also be expressed similarly as $\Tau^\mathscr{F} = N \cdot \tau^{\mathscr{F}}$, $\Tau^\mathscr{R} = N \cdot \tau^{\mathscr{R}}$ and $\Tau^\mathscr{P} = N \cdot \tau^{\mathscr{P}}$ respectively, where $\tau$ is the walltime taken by each corresponding operation. The total walltime can thus be written as
\begin{gather}
\Tau^\text{\AlgAbb} = K \cdot \Big[ \tau^\text{f} + N \cdot ( \tau^\text{c} + \tau^{\mathscr{F}} + \tau^{\mathscr{R}} + \tau^{\mathscr{P}} ) \Big] .
\label{eqn:parareal}
\end{gather}
In contrast, the same simulation with only the fine solver running in serial would take $\Tau^\text{serial} = N \cdot \tau^\text{f}$. When conventional domain decomposition is applied using the same number of cores, the walltime is reduced to
\begin{equation}
\Tau^{\text{serial}}_\text{SD} = \beta \cdot N \cdot \tau^{f},
\label{eqn:serial}
\end{equation}
where $N^{-1} \leq \beta \leq 1$. For methods that scale perfectly linearly, $\beta$ is equal to $N^{-1}$. In the case that additional resources do not result in speedup, $\beta$ is equal to $1$.
Comparing the walltime of \AlgAbb~in Eq.~\eqref{eqn:parareal} with conventional domain decomposition in Eq.~\eqref{eqn:serial}, \AlgAbb~offers better speedup when
\begin{equation}
\beta > \dfrac{K \cdot \Big[ \tau^\text{f} + N \cdot ( \tau^\text{c} + \tau^{\mathscr{F}} + \tau^{\mathscr{R}} + \tau^{\mathscr{P}} ) \Big]}{N \cdot \tau^\text{f}}.
\label{eqn:comp}
\end{equation}
Typically, $\tau^\text{f} >> \tau^\text{c}$, and both $\tau^\text{f}$ and $\tau^\text{c}$ are much larger than $\tau^{\mathscr{F}}$, $\tau^{\mathscr{R}}$ and $\tau^{\mathscr{P}}$. Therefore, Eq.~\eqref{eqn:comp} can be reduced to
\begin{gather}\label{eq:beta}
\beta \gtrapprox K \Big[ \dfrac{1}{N} + \dfrac{\tau^\text{c}}{\tau^\text{f}} \Big] .
\end{gather}

\begin{figure}[t!]
	\centering
	\includegraphics[width=0.5\textwidth]{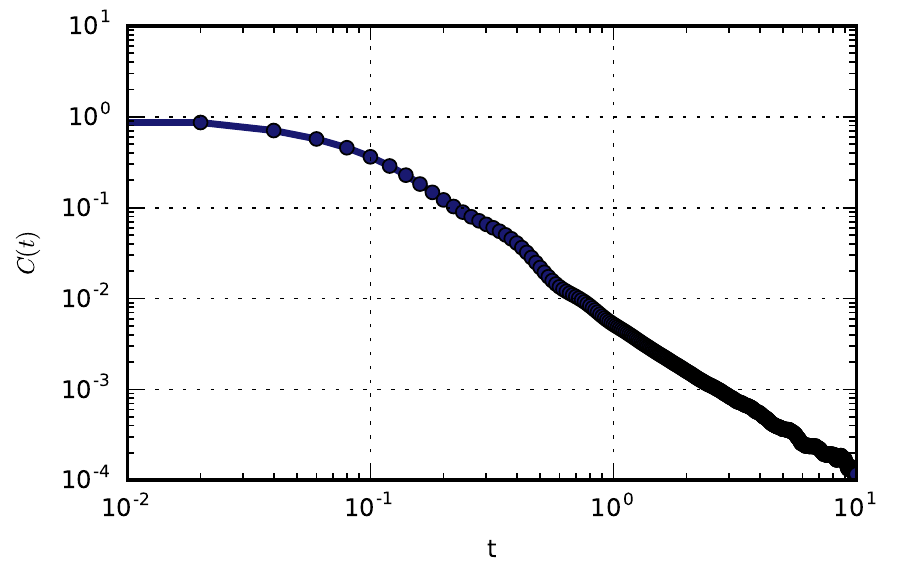}
	\caption{Normalized temporal correlation function of momentum, $C(t)=\langle \mathbf{p}(t)\mathbf{p}(0) \rangle$, computed from a simple DPD fluid is plotted against time. The decay of $C(t)$ signifies momentum decorrelation over time for simple isothermal DPD fluid. From our benchmark tests, $C(t)=10^{-4}$ provides a sufficient decorrelation of momentum and leads to accurate results. Therefore, in the present work, we use $\Delta T=10$, where the value of $C(t)$ is approximately $10^{-4}$.}
    \label{fig:acf}
\end{figure}

We note that in order to select the number of time subdomains, we must carefully consider the physics of the underlying stochastic dynamics. In particular, each particle model is associated with a set of dynamical properties, which describe the way collective behavior leads to macroscopic observables. For hydrodynamics, this is given by the characteristic time for momentum decorrelation, which can be quantified by the temporal correlation function of momentum, defined as $C(t)=\langle \mathbf{p}(t+\tau)\mathbf{p}(\tau) \rangle$, where $\mathbf{p}(t)=mass \cdot \mathbf{v}(t)$ represents the instantaneous momentum. The computed $C(t)$ of simple isothermal DPD fluid is shown in Fig.~\ref{fig:acf}, normalized by the value of $C(0)$. The decay of $C(t)$ signifies momentum decorrelation over time, and correspondingly we should use a suitable $\Delta T$ in our Parareal algorithm so that momentum is sufficiently decorrelated. From our benchmark tests, $C(t)=10^{-4}$ provides a sufficient decorrelation of momentum and leads to accurate results. Therefore, in the present work, we use $\Delta T=10$, where the value of $C(t)$ is approximately $10^{-4}$ according to Fig.~\ref{fig:acf}.

\subsection{Methods}
\AlgAbb~offers great versatility and robustness in terms of propagator choices. As a framework targeting particle methods, \AlgAbb~does not limit the type of particle solvers; that is to say, any particle solver can be accelerated with \AlgAbb. For demonstration, we model mesoscale hydrodynamics with DPD, a stochastic Lagrangian method. DPD can be derived rigorously from coarse-graining of atomistic dynamics~\cite{2014ZLi_SM, 2015ZLi_JCP} and can recover macroscopic hydrodynamics in the continuum limit~\cite{2017ZLi_chapter}. For a simple DPD fluid, a suitable low-dimensional model would be the continuum representation of fluid, namely the Navier-Stokes equations. Since there is no constraint on the discretization of the continuum equations, for simplicity, we chose the finite difference (FD) scheme as the coarse propagator.

The concurrent coupling of two independent solvers requires robust on-the-fly data transfer. For the examples shown here, we used a light weight library called Multiscale Universal Interface~\cite{tang2015multiscale}, or MUI for short, to bridge the solvers.

\subsubsection{Dissipative Particle Dynamics}
In the DPD framework, each particle is represented explicitly by its position and velocity. The time evolution of DPD particles is governed by Newton's equation of motion~\cite{Hoogerbrugge1992}:
\begin{align} \label{eqn:Newton}
& \frac{d\mathbf{r}_i}{dr} = v_i ,\\
& \frac{d\mathbf{v}_i}{dt} = \mathbf{F}_i = \sum_{i \neq j} (\mathbf{F}^C_{ij} + \mathbf{F}^D_{ij} + \mathbf{F}^R_{ij}),
\end{align}
where $t$, $\mathbf{r}_i$, $\mathbf{v}_i$ and $\mathbf{F}_i$ denote time, position, velocity, and force, respectively. The force imposed on particle $i$ is the sum of conservative force $\mathbf{F}^C_{ij}$, dissipative force $\mathbf{F}^D_{ij}$, and corresponding random force $\mathbf{F}^R_{ij}$ from particle $j$ within a radial cutoff $r_c$ of $i$. Those pairwise forces are expressed as~\cite{Groot1997b}:
\begin{align} \label{DPDforce}
& \mathbf{F}^C_{ij} = \alpha_{ij} \; \omega_C(r_{ij}) \; \mathbf{e}_{ij}, \\
& \mathbf{F}^D_{ij} = -\gamma_{ij} \; \omega_D(r_{ij}) \; ( \mathbf{e}_{ij} \cdot \mathbf{v}_{ij} ) \; \mathbf{e}_{ij}, \\
& \mathbf{F}^R_{ij} = \sigma_{ij} \; \omega_R(r_{ij}) \; \xi_{ij} \; \Delta t_f^{-1/2} \; \mathbf{e}_{ij},
\end{align}
where $\mathbf{e}_{ij}=\mathbf{r}_{ij} / r_{ij}$ is the unit vector between particles $i$ and $j$, and $\mathbf{v}_{ij}$ is the velocity difference. $\Delta t_f$ is the time step of the fine propagator, and $\xi$ is a symmetric Gaussian random variable with zero mean and unit variance~\cite{Groot1997b}. The model parameters $\alpha_{ij}$, $\gamma_{ij}$ and $\sigma_{ij}$ adjust the conservative, dissipative, and random forces, respectively. The corresponding weighting function $\omega_C(r_{ij})$, $\omega_D(r_{ij})$ and $\omega_R(r_{ij})$ regulate the interaction between pairs of particles. The balance between dissipation and thermal fluctuation is maintained by constraints from the fluctuation-dissipation theorem~\cite{Espanol1995a}:
\begin{equation} \label{FDTforce}
\sigma^2_{ij} = 2 \gamma_{ij} k_B T , \;\;\;\;\;\; \omega_D(r_{ij}) = \omega_R^2(r_{ij}),
\end{equation}
where $k_B T$ is the Boltzmann energy unit. A common choice for the weight function is $\omega_C(r)=1-r/r_c$ and $\omega_D(r)=\omega^2_R(r)=(1-r/r_c)^2$ for $r\leq r_c$ and zero for $r>r_c$, which is used for the simulations shown in this paper. For simplicity, $k_BT$ and the mass of a particle are taken as the energy unit and mass unit, and their values are set to one.

\subsubsection{Model Coupling}
By applying a Mori-Zwanzig projection operator technique to a DPD system, Espa{\~n}ol~\cite{1995Espanol} and Marsh et al.~\cite{1997Marsh} formally derived the hydrodynamic equations of a simple DPD fluid using the Green-Kubo formulas, recovering the continuity and Navier-Stokes equations. For more details on the mathematical derivations leading to macroscopic hydrodynamic equation from DPD dynamics, we refer the interested readers to a paper by Espa{\~n}ol~\cite{1995Espanol} and a book by Zwanzig~\cite{2001Zwanzig}. Therefore, the continuity and Navier-Stokes equations can be considered as a mean-field representation of the DPD system, and consequently provide macroscopic solution as a rough prediction of the hydrodynamics that can be used to supervise the DPD simulations in the time domain.

For coupling the grid-based Navier-Stokes solver with the Lagrangian DPD solver, we found MUI easy to implement in the sense that codes do not need to be refactored before application. Adding only a few lines of code to existing solvers is sufficient to adhere to MUI's push-fetch workflow - data pushed by one solver is then fetched by the other. In the context of particle solvers, the relevant data are particle positions and the associated quantities. For continuum solvers, the quantities of interest are associated with nodes of the discretization grid. In a data-transfer, the sender pushes the data into a MUI Interface Layer (\textbf{IF}), which then assists the receiver to fetch and decode these data based on the topological configuration. Because topological configuration in the receiver usually do not exactly overlap with these of the particle solver, a sampler such as Gaussian kernel or nearest neighbor might be used to calculate an appropriate nodal value in the receiver. As for the linking configuration, each of the fine propagators is attached to one \textbf{IF}, and all of those \textbf{IF}s link to the coarse propagator.

\section{Numerical Results}\label{sec:3}
In our proposed framework, the macroscopic model is a mean-field approximation to our microscopic model. As a consequence, the physical parameters in the macroscopic model cannot be calculated or deduced precisely from the microscopic model. However, the macroscopic model in \AlgAbb~only requires rough estimations of these physical parameters, effects of which can be iteratively “corrected” by the correction term in Algorithm~\ref{alg:MultiParareal}. In the example problems below, we also demonstrate empirically that a more accurate parameter estimation in the macroscopic model leads to faster convergence. In addition, we will analyze the accuracy and convergence of \AlgAbb~on the example problems.

\subsection{One-dimensional time-dependent flow} \label{Exproblem}
As a demonstration, we applied \AlgAbb~to the one-dimensional time-dependent plane Poiseuille flow with DPD and FD as fine and coarse propagators, respectively. Simulating Poiseuille flow is a well-accepted benchmark test for many Lagrangian methods including DPD because the macroscopic parameter viscosity can be computed empirically from simulations~\cite{backer2005poiseuille}. We refer to the computed viscosity as the true viscosity $\nu_\text{true}$, which is incorporated in the true solution that is used in error analysis.

Macroscopically, the pressure-driven Poiseuille flow can be described by the incompressible Navier-Stokes equations:
\begin{gather}
\frac{\text{D} \mathbf{u}}{\text{D} t} = \nu \nabla^2 \mathbf{u} + F,
\end{gather}
where $\nu$ is the viscosity, $F$ denotes a body-force driven by a constant pressure gradient, and $\mathbf{u}$ denotes the velocity field. The exact time-dependent solution to the Navier-Stokes equation is known analytically~\cite{sigalotti2003sph}:
\begin{equation}
u(x,t)=\frac{Fd^2}{8\nu}\Big(1-\Big(\frac{2x}{d}^2\Big)\Big) - \sum^\infty_{n=0} \frac{4(-1)^n Fd^2}{\nu \pi^3(2n+1)^3} \cdot \cos\Big(\frac{(2n+1) \pi x}{d} \Big) \cdot \exp \Big(-\frac{(2n+1)^2 \pi^2 \nu t}{d^2} \Big),
\label{eqn:poise_soln}
\end{equation}
The true solution $u(\nu_\text{true})$ can thus be readily calculated according to Eq.~\eqref{eqn:poise_soln}. For this problem, we use a periodic box with the size of $30 \times 20 \times 10$ in DPD units containing $24,000$ particles. The DPD parameters $\alpha$ and $\gamma$ are set to $18.75$ and $4.5$, respectively. All particles are subject to a body-force $F=0.1$ and a radial cutoff $r_c=1.58$. Given these parameters, the true viscosity for this system is approximately $0.841$ computed via DPD simulations.

As explained in Section~\ref{subsec:TheoSpeedup}, the momentum is sufficiently decorrelated beyond $10$ time units according to the temporal correlation function $C(t)$. The communication interval between the propagators $\Delta T$, also known as the time step of the coarse propagator, is thus set to be $10$. With the time step $\Delta t_f=0.01$, the fine propagators march forward for $1000$ time steps during the same period. Because the Poiseuille flow is one-dimensional, we use a simple linear interpolation as the projection operation. To find the velocity profile, the simulation domain is divided into slabs of uniform width that are also the size of spatial discretization in the coarse propagator. We then average particle velocities in each slab. The result is a one-dimensional velocity profile that describes the macroscopic state. On the other hand, the macroscopic state is mapped to the particle system with a nearest-neighbor sampler - a particle takes on the velocity of the nearest grid node. The coarse propagation and the filtering of stochastic noise generated by the fine propagator are carried out simultaneously in this example. The viscosity of fluid can naturally create a low-pass filtering effect~\cite{2009Hasegawa} that damps high-frequency fluctuations as the coarse solver advances. To take advantage of the filtering effect, one time subdomain $\Delta T=10$ is covered by $100$ sequential coarse steps with an effective time step $\Delta t_c = 0.1$.

\subsubsection{Impact of Filtering} \label{sec:ImpactOfFiltering}
As an iterative scheme, the convergence of \AlgAbb~greatly relies on the notion that, for a propagation, the stochastic noise in the model should not cause the solution to deviate far from the ensemble solution. To ensure convergence, a filtering operation must be incorporated in the algorithm. If the filtering operation is not enforced, the noise can grow in the temporal space within an iteration and will propagate across iterations. Unchecked noise growth can eventually drive the propagated solution away from the true solution. To demonstrate the impact of filtering, we compared results of the Poiseuille flow simulations without and with partial filtering. For fluids, the macroscopic model has a filtering mechanism build-in: the viscosity of fluid is a natural low-pass filter that damps high-frequency fluctuations. To take advantage of the filtering effect, a time subdomain $\Delta T$ is divided into smaller (effective) time steps $\Delta t_c$ which the coarse solver uses.

\begin{figure}[h!]
	\centering
    \begin{tabular}{c c}
	\subfloat[]{\includegraphics[width=0.5\textwidth]{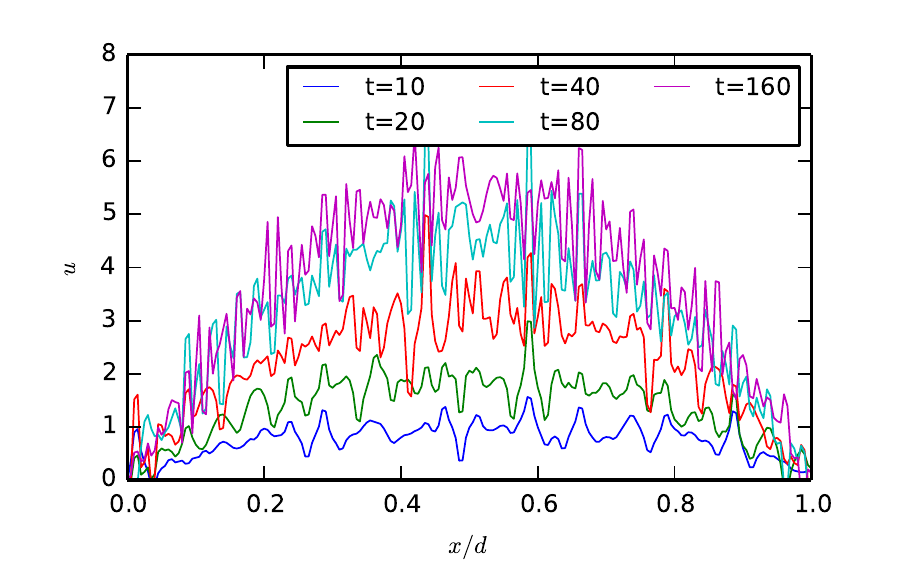}\label{wofp}}
	\subfloat[]{\includegraphics[width=0.5\textwidth]{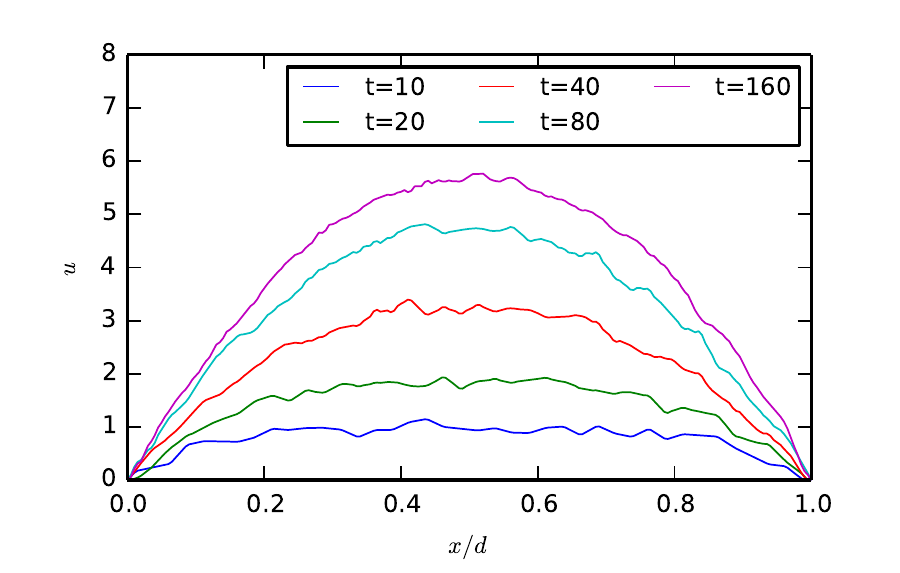}\label{wfp}} \\
    \subfloat[]{\includegraphics[width=0.5\textwidth]{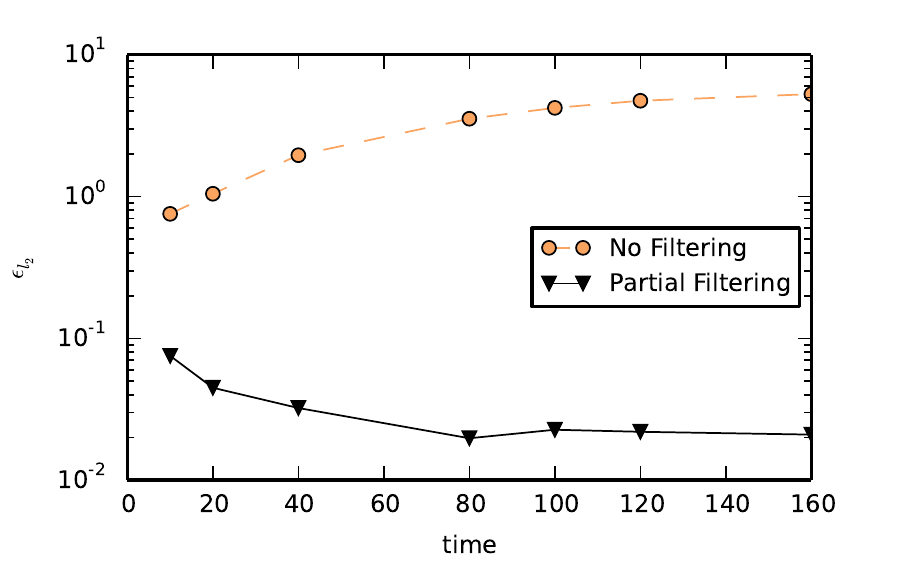}\label{fl2time}}
    \subfloat[]{\includegraphics[width=0.5\textwidth]{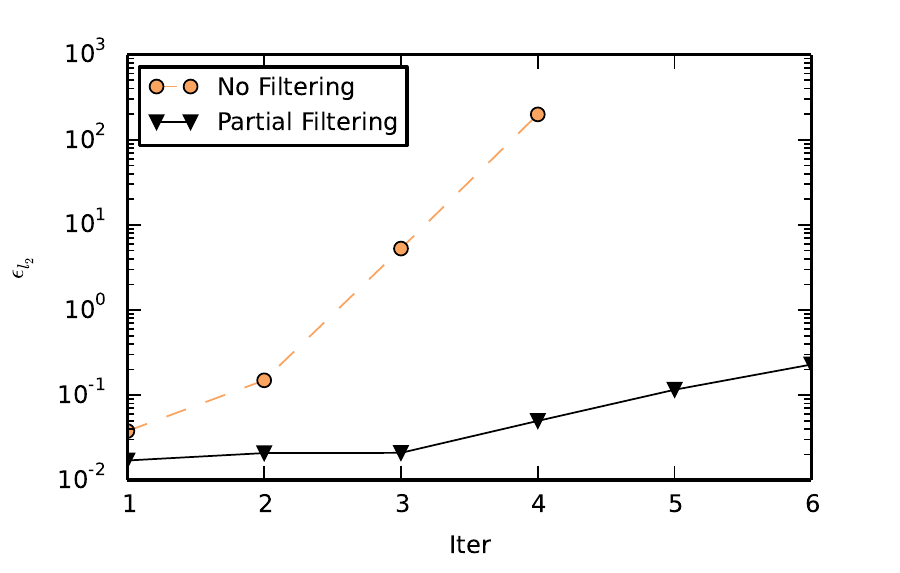}\label{fl2}}
	\end{tabular}
	\caption{The impact of filtering is apparent when the simulations (a) without and (b) with partial filtering are compared. In the partial-filtering case where the length of a time subdomain is $10$ units, the coarse propagator with an effective time step of $1$ unit filters the solution 10 consecutive times. The noise is much more prominent when the length of a time subdomain is reduced to $1$ unit while the effective time step is unchanged. In that case, filtering is not performed. (c) Moreover, quantified as normalized $l_2$ error, the noise increases monotonically with time within an iteration. (d) Across iterations, it also builds up and eventually overtakes the solution.}
	\label{fig:W_WO_filtering}	
\end{figure}

To demonstrate the impact of filtering, we run a simple example without and with partial-filtering. The same effective time step, subsequently the same Courant-Friedrichs-Lewy condition, are used in both simulations to isolate the effect of filtering. For a time subdomains of $10$ time units, the coarse propagator with an effective time step of $1$ unit filters the solution 10 consecutive times. This approach offers better noise reduction than the case with a subdomain size of $1$ unit and an effective time step of $1$ unit, which offers no filtering. The impact of filtering is apparent in this comparison: the solution in the no-filtering case in Fig.~\ref{fig:W_WO_filtering}\protect\subref{wofp} is substantially nosier than the one in the partial-filtering case in Fig.~\ref{fig:W_WO_filtering}\protect\subref{wfp}. To quantify the deviation from the reference solution, we compute the normalized $l_2$ error, expressed as
\begin{gather}
\epsilon_{l_2} \equiv \frac{ \Big\lVert u^\text{ref} - u^\text{sim} \Big\rVert_{l_2} } {\Big\lVert u^\text{ref} \Big\rVert_{l_2}} , \\
\Big\lVert u^\text{ref} - u^\text{sim} \Big\rVert_{l_2} = \Big( \sum_i \abs{u_i^\text{ref}-u_i^\text{sim}}^2 \cdot \Delta x \Big)^{1/2},
\end{gather}
where $\Delta x$ denotes the size of spatial discretization, and $i$ is the nodal index. As shown in Fig.~\ref{fig:W_WO_filtering}\protect\subref{fl2time}, the normalized $l_2$ error $\epsilon_{l_2}$ for the no-filtering case is monotonically increasing with time, which signifies the growth of noise in time within an iteration. The noise also propagates across iterations shown in Fig.~\ref{fig:W_WO_filtering}\protect\subref{fl2}. For the partial-filtering case, the noise accumulates at a much slower rate than the no-filtering case.

\subsubsection{Accuracy}
\begin{figure}[h!]
	\centering
    \begin{tabular}{c c}
	\subfloat[]{\includegraphics[width=0.5\textwidth]{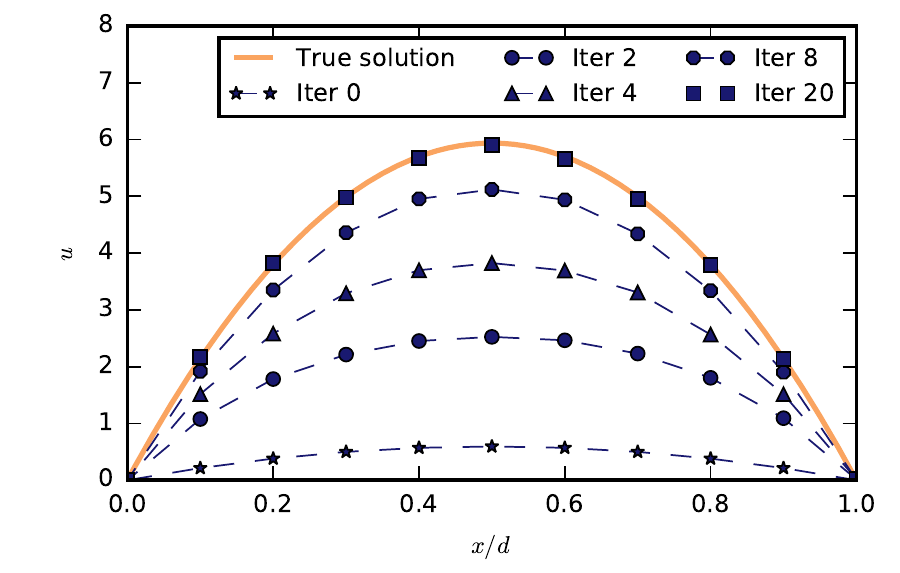}\label{iterevo}}
	\subfloat[]{\includegraphics[width=0.5\textwidth]{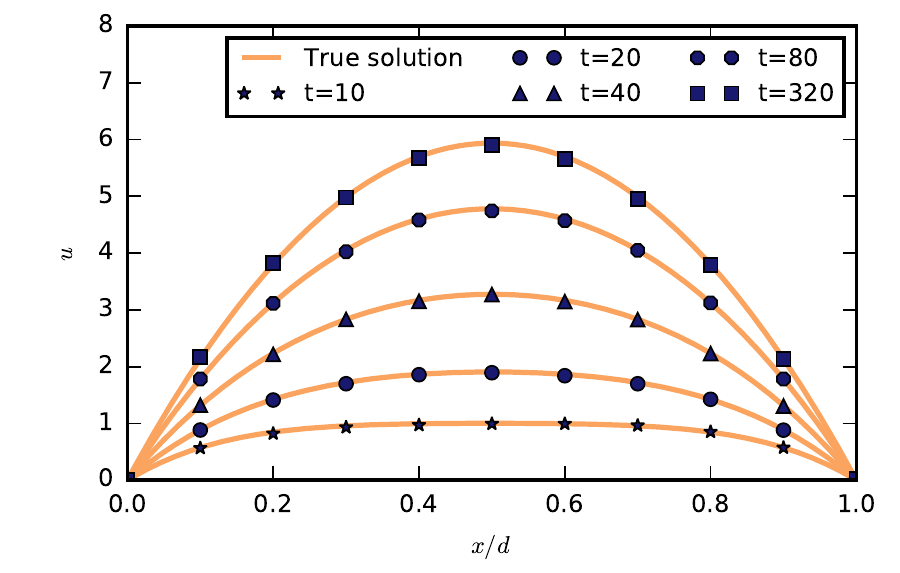}\label{timeevo}}
	\end{tabular}
	\caption{Results of plane Poiseuille flow simulation with \AlgAbb. (a) Evolution of steady-state velocity profile over iterations. The parameter viscosity in the low-dimensional model is $10\times$ the true value. In that case, the velocity profile obtained with \AlgAbb~is refined and converges to the true solution iteratively. (b) Transient velocity profiles obtained with \AlgAbb. On Iteration $20$, the profiles obtained with \AlgAbb~are compared with the true velocity profiles. This demonstrates that \AlgAbb~is able to obtain correct transient solutions as well as the steady state solution. }
	\label{fig:velocity_profile}	
\end{figure}

\begin{figure}[h!]
	\centering
	\begin{tabular}{c c}
	\subfloat[]{\includegraphics[width=0.5\textwidth]{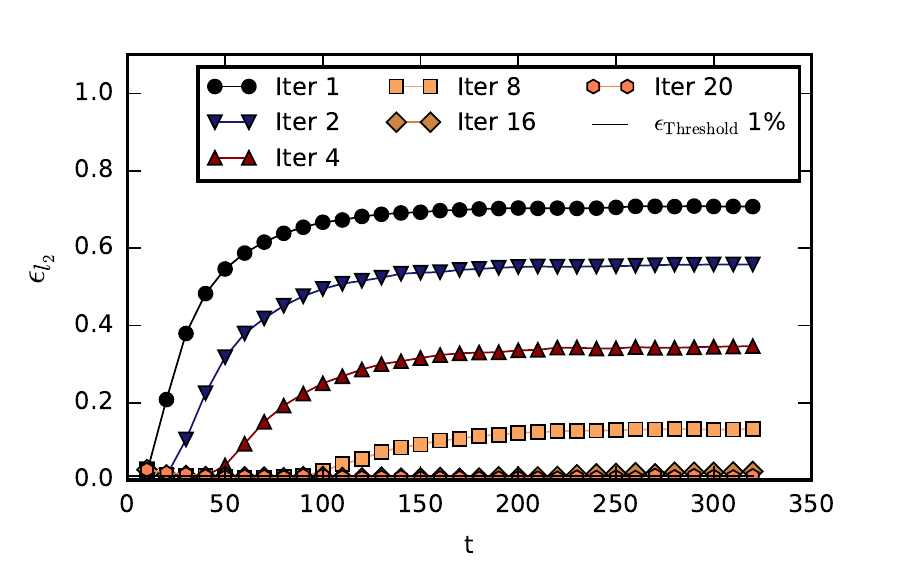}\label{fig:Poise_l2norm}} &
	\subfloat[]{\includegraphics[width=0.5\textwidth]{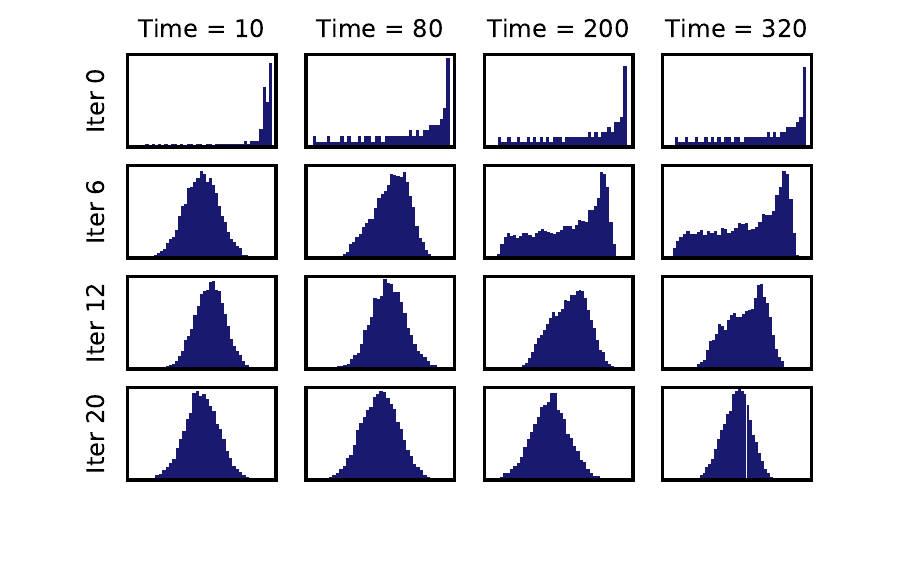}\label{fig:Poise_distri}}
	\end{tabular}
	\caption{Analyzing the results of plane Poiseuille flow simulation with \AlgAbb. (a) Normalized $l_2$ error between simulation and true solution, denoted as $\epsilon_{l_2}$. Because of intrinsic noise from the microscopic model, we accept solutions within $1\%$ difference. (b) Normalized $l_2$ error histogram. The error histogram eventually recovers Gaussian distribution for all times, indicating convergence. }
	\label{fig:Poise}
\end{figure}

To demonstrate the flexibility and robustness of \AlgAbb, we purposely estimated the viscosity in the macroscopic model to be $10\times$ the true viscosity. In other words, the estimated mean-field solution $u(\nu_{\text{est}}) = u(\nu_{\text{true}}) / 10$.
To verify the accuracy of \AlgAbb, we compare the true velocity profiles with the ensemble average obtained with multiple \AlgAbb~runs. Because the microscopic model is stochastic in nature, it is desirable to obtain an ensemble averaged solution that is less distorted by stochastic noise. In our experiment, the simulated velocity profile is ensemble-averaged from fifty independent simulations. The resulting transient velocity profiles, as well as the evolution of velocity profile at steady state, match those of the true velocity profiles visually as shown in Fig.~\ref{fig:velocity_profile}. To quantify the accuracy of \AlgAbb~, we computed the normalized $l_2$ error $\epsilon_{l_2}$ between the reference and simulated solutions. For the one-dimensional Poiseuille plane-flow, the reference solution is the true solution $u(\nu_\text{true})$ calculated according to Eq.~\eqref{eqn:poise_soln} because the analytical form is known. We found that $\epsilon_{l_2}$ decreases over iterations as shown in Fig.~\ref{fig:Poise}\protect\subref{fig:Poise_l2norm}. Given that the noise cannot be eliminated by averaging fifty solutions, we accept solutions within $1\%$ error as a threshold denoted as $\epsilon_{\text{Threshold}}$. In this particular example, $\epsilon_{l_2}$ falls to $\epsilon_{\text{Threshold}}$ in $20$ iterations. At the threshold, the error has a Gaussian distribution as shown in Fig.~\ref{fig:Poise}\protect\subref{fig:Poise_distri}, which matches the noise distribution from the microscopic model~\cite{Groot1997b}. The recovery of the distribution signifies that the \AlgAbb~preserved stochastic fluctuations of the microscopic model.

\subsection{Convergence}
\begin{figure}[h!]
	\centering
	\begin{tabular}{c c}
	\subfloat[]{\includegraphics[width=0.5\textwidth]{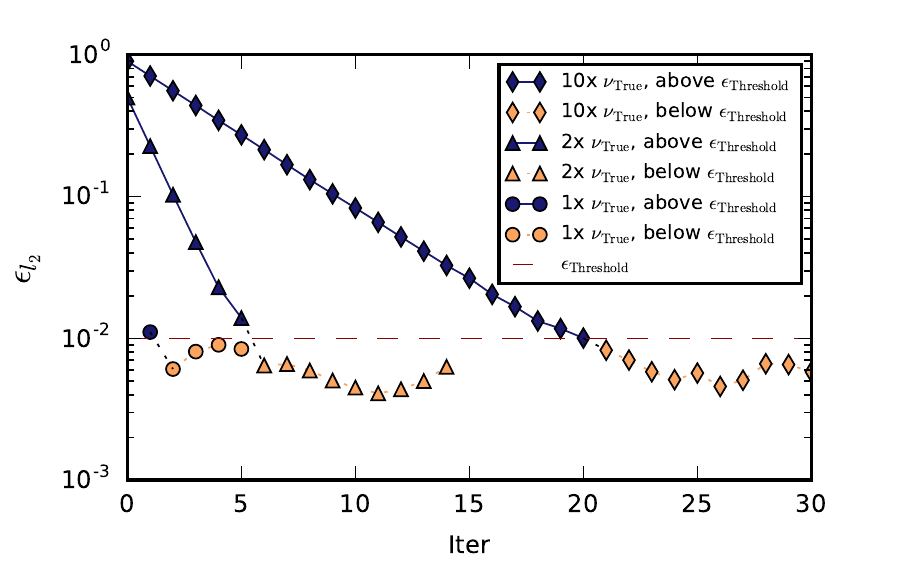}\label{fig:rateofconv}} &
	\subfloat[]{\includegraphics[width=0.5\textwidth]{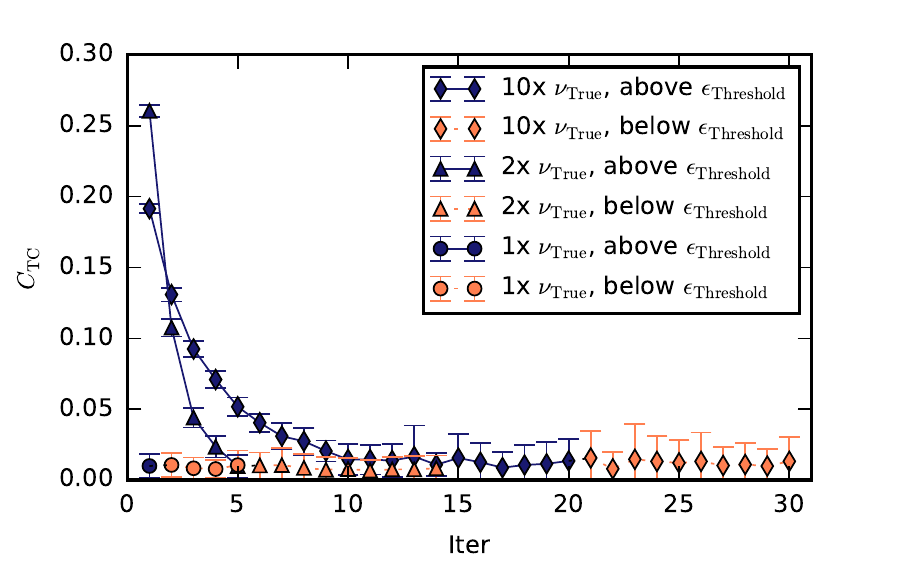}\label{fig:convthreshold}}
	\end{tabular}
	\caption{(a) l2 error. $\epsilon_{l_2}$ at $T_\text{final}$ is plotted against iterations. In addition to $10\times$, two experiments were performed with estimated viscosities that are $2\times$ and $1\times$ true viscosity. The nearly linear plot on a semi-log scale indicates exponential convergence. Moreover, when true viscosity is used, it takes one iteration to reach the error threshold. (b) Termination condition. The termination criterion $C_\text{TC}$ from Eq.~\eqref{eqn:term_condition} is plotted. Solid navy lines correspond to iterations before the error threshold, and dotted yellow lines after the error threshold.}
	\label{fig:Poise_conv}
\end{figure}

Inspecting $\epsilon_{l_2}$ at $T_\text{final}=320$ for all iterations in Fig.~\ref{fig:Poise_conv}\protect\subref{fig:rateofconv}, \AlgAbb~achieves exponential convergence. We repeat the experiment with a more accurate macroscopic viscosity at $2\times$ true viscosity and obtain exponential convergence as well, but only $5$ iterations are needed to reach the same threshold. More importantly, $\epsilon_{l_2}$ reaches a minimum level regardless of the estimated viscosity. The minimum error is determined by many factors; in addition to the error intrinsic to Parareal, operations in \AlgAbb~such as filtering and projection introduce systematic errors that are difficult to separately quantify.

Lastly, we calculate $C_\text{TC}$ from Eq.~\eqref{eqn:term_condition} and plot it in Fig.~\ref{fig:Poise_conv}\protect\subref{fig:convthreshold}. Because of the stochastic noise, $C_\text{TC}$ varies between runs. We plot the mean and $95\%$ confidence interval for $C_\text{TC}$ generated from fifty independent runs. When the parameter estimation is more accurate as in the case of 2x true viscosity, $C_{\text{TC}}$ fluctuates mildly as indicated by the range of the confidence interval and drops sharply to a level that satisfies the termination condition. However, when estimation is rough, the confidence interval is relatively wide without a clear boundary separating iterations above and below $\epsilon_{\text{Threshold}}$. The resulting consequence is discussed in Section~\ref{sec:SumDis}.

\subsection{Efficiency}
For particle simulations, spatial domain decomposition is the most utilized acceleration method because the subdomains can be easily assigned to different processors. To compare the performance of \AlgAbb~with conventional spatial decomposition (CSD), we apply them separately to the planer Poiseuille flow problem and then calculate the parallel efficiency.

The same simulation parameters described in Section~\ref{Exproblem} are used in this benchmark test. To avoid implementing no-slip boundaries, we simulated a reverse Poiseuille flow~\cite{2005Backer,2014ZLi_JCP}, in which two planar flows are stacked to create periodic boundaries in all directions. At the same time, the $z$-direction is shrunk by one half to preserve the total number of particles. For a general comparison, we simulate a single iteration, and each core was responsible for $10$ time units. When the number of available cores is doubled, the final time $T_{\text{final}}$ is also doubled.

\begin{figure}[t!]
	\centering
	\begin{tabular}{c c}
	\subfloat[]{\includegraphics[width=0.5\textwidth]{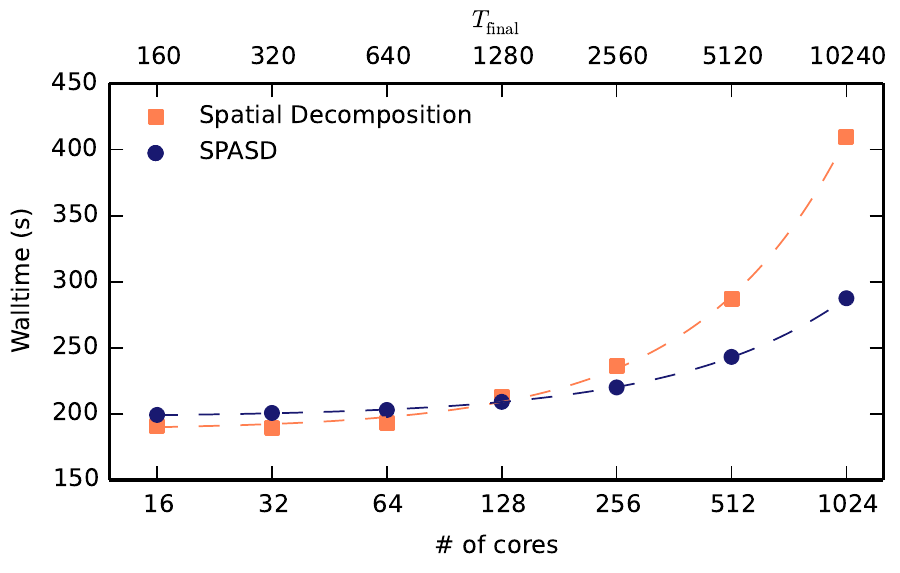}\label{Walltime}}
	\subfloat[]{\includegraphics[width=0.5\textwidth]{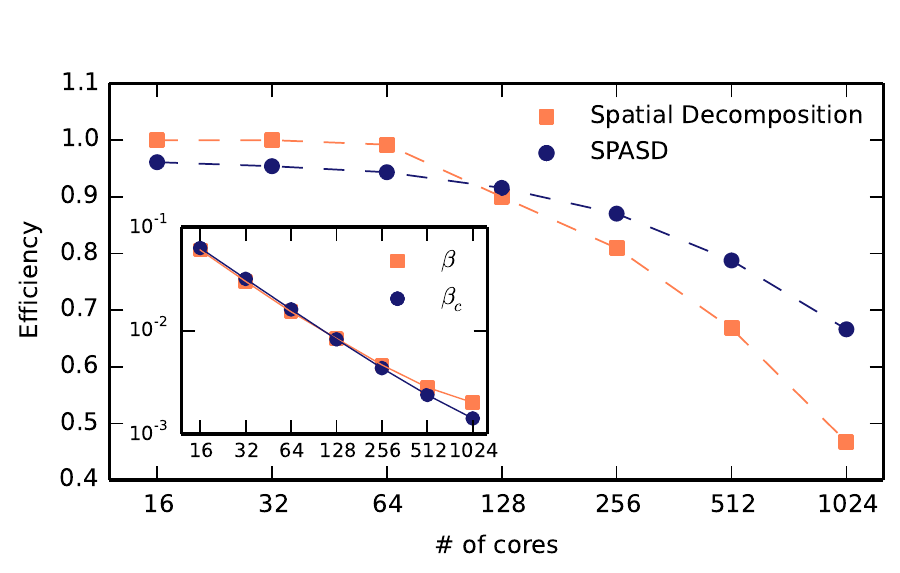}\label{Efficiency}}
	\end{tabular}
	\caption{Comparison between \AlgAbb~and conventional spatial decomposition (CSD). (a) Walltimes of simulation with CSD and simulation with \AlgAbb. The performances of CSD and \AlgAbb~are similar when fewer cores are available. As the core-count and $T_\text{final}$ simultaneously increase, the performance of \AlgAbb~starts to dominate. The walltime of CSD grows exponentially at a much faster rate than \AlgAbb. (b) Using the walltime of a single-core simulation at different $T_\text{final}$ as a basis, the speedup of CSD and \AlgAbb~can be computed. The efficiencies were then calculated as the speedup per core. The higher parallel efficiency at high core-counts indicates that \AlgAbb~is a more scalable algorithm for long-time simulations. In reference to Eq.~\eqref{eqn:comp}, we also compared $\beta = \Tau^\text{serial}_\text{SD} / (\tau^f N)$ and $\beta_c = [ \tau^\text{f} + N \cdot ( \tau^\text{c} + \tau^{\mathscr{F}} + \tau^{\mathscr{R}} + \tau^{\mathscr{P}} ) ] / (N \cdot \tau^f)$. Starting from $128$ cores and up, $\beta$ begins to flatten out while $\beta_c$ continues to drop. The trend echoes with the efficiency plot and reaffirms that \AlgAbb~starts to dominate as few as $128$ cores for this example.}
	\label{fig:Efficiency}
\end{figure}

This particular benchmark was run on nodes with AMD Opteron 6274 CPUs. For this example, the performances of CSD and \AlgAbb~are similar when fewer cores are available as shown in Fig.~\ref{fig:Efficiency}\protect\subref{Walltime}. As the core-count and $T_\text{final}$ increase, their performances start to diverge at approximately $128$ cores. From there, the walltime of CSD grows at a much faster exponential rate than that of \AlgAbb. Two factors contribute to this: First, the size of a spatial subdomain in CSD is limited by the average particle distance. Second, the time spent on communication between nodes grows at a rate that nullifies the benefits of additional cores. \AlgAbb, however, does not suffer from the same issues. Communication between nodes is only invoked at the end of temporal subdomain, and the duration of temporal subdomain can be adjusted at will. This makes \AlgAbb~a more scalable algorithm suitable for long-time simulations. For comparison, we used the speed of a single-core simulation, roughly $39$ seconds per time unit, as a basis and computed the speedup of each method. The efficiencies were then calculated as the speedups per core. As expected, \AlgAbb~offers better parallel efficiency in the cases of large core counts, as shown in Fig.~\ref{fig:Efficiency}\protect\subref{Efficiency}.

The walltime taken by the coarse propagator and associated operations $\tau^\text{c} + \tau^{\mathscr{F}} + \tau^{\mathscr{R}} + \tau^{\mathscr{P}}$, as well as the walltime taken by the fine propagator $\tau^\text{f}$, can be readily calculated with a simple linear fit to Eq.~\eqref{eqn:parareal}. For this benchmark case, the walltime of the simulation using \AlgAbb~is described by $\Tau^\text{\AlgAbb} = 197.8 + 0.08783 N$ where $N=T_\text{final}/10$. Plugging in the numerical value for $\tau^\text{f}$, the walltime of simulation using CSD is reduced to $\Tau^{\text{serial}}_\text{SD} = 197.8 \beta N$. The scaling factor $\beta$ depends on many factors including the number of parallel cores, distribution topology, and parallel-method efficiency. When few cores are available, $\beta$ is approximately equal to $N^{-1}$ as shown in Fig.~\ref{fig:Efficiency}\protect\subref{Efficiency}, and CSD provides better efficiency and performance. As the number of cores increases, $\beta$ decreases proportionally until around 128 cores at which \AlgAbb~and CSD have similar parallel efficiencies. At this core-count, $\beta$ was calculated to be $0.0083$, which agrees with our prediction $\beta_c \equiv K(N^{-1}+\tau_c/\tau_f) = 0.0084$ given by Eq.~\eqref{eq:beta}. From this core-count forward, the flattening of the $\beta$ curve signifies the drop of CSD efficiency.

\subsection{Two-dimensional time-dependent flow}
To further demonstrate the applicability of \AlgAbb, we simulated a two-dimensional cavity flow. Simulating transient cavity flow is much more complicated than the Poiseuille flow because the convective acceleration and the corner singularity induced by the cavity make it numerically challenging. However, the singularities, including corner singularity~\cite{2006Nie}, crack propagation~\cite{2014Bouchbinder}, and moving contact line singularity~\cite{2013Sibley}, in partial differential equation (PDE) models (in the limit of $\Delta x\rightarrow0$) represent rich underlying multiscale physics, and can be captured by particle solvers, e.g., DPD or MD.

In this case, we again use DPD to simulate the lid-driven cavity flow, which is supervised in the temporal domain by the incompressible Navier-Stokes equations.
These equations are solved with finite difference on a staggered grid: the pressure is stored at the center of the grid cells, and the velocities are stored at the faces of the grid cells. The second order spatial derivative is approximated with a centered difference scheme. This setup allows pressure and velocity to be solved concurrently in a two-step process. The first step is to compute an intermediate velocity by solving the momentum equation, omitting the effect of pressure. The intermediate velocity is then used to compute new pressure by solving a pressure Poisson equation. The second step is to solve for the new velocity also using the intermediate velocity and pressure. For temporal discretization, we use explicit time stepping.

\begin{figure}[t!]
	\centering
	\includegraphics[width=1.0\textwidth]{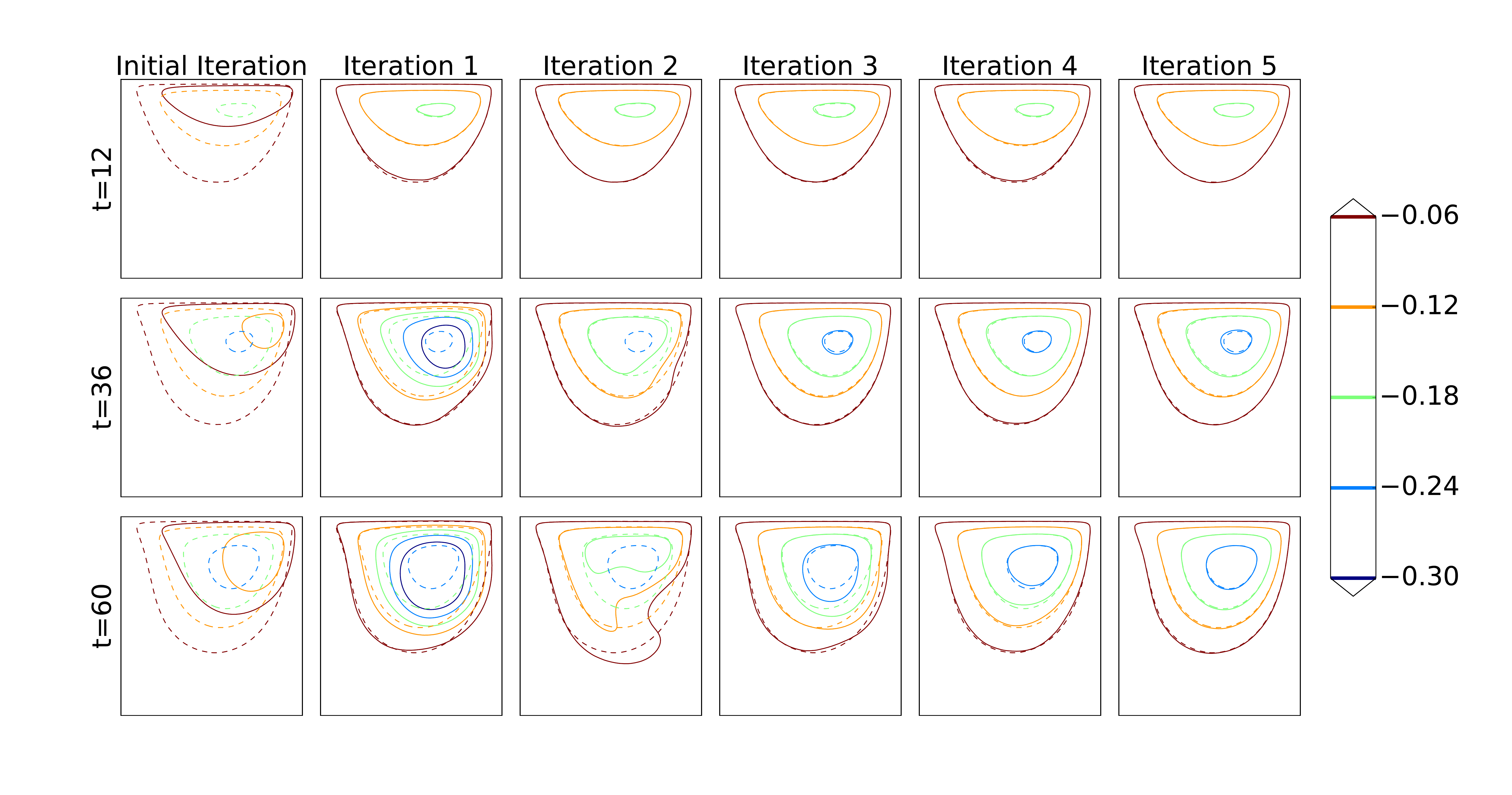}
	\caption{Stream function of the two-dimensional cavity flow with \AlgAbb. The dashed and solid contour lines represent the reference and simulation solutions, respectively. In the initial iteration, the center of the vortex matches poorly with the reference solution. This is expected because the low-dimensional model is solved with an estimated Reynolds number $\text{Re}_\text{est}$ that is $3\times$ the reference Reynolds number $\text{Re}_\text{ref}$, and only the predictor is involved computing the flow in the initial iteration. On the third iteration, the center of the vortex converges to the reference vortex.}
	\label{fig:cavity_diff}
\end{figure}

\begin{figure}[h!]
	\centering
	\begin{tabular}{c c}
	\subfloat[]{\includegraphics[width=0.5\textwidth]{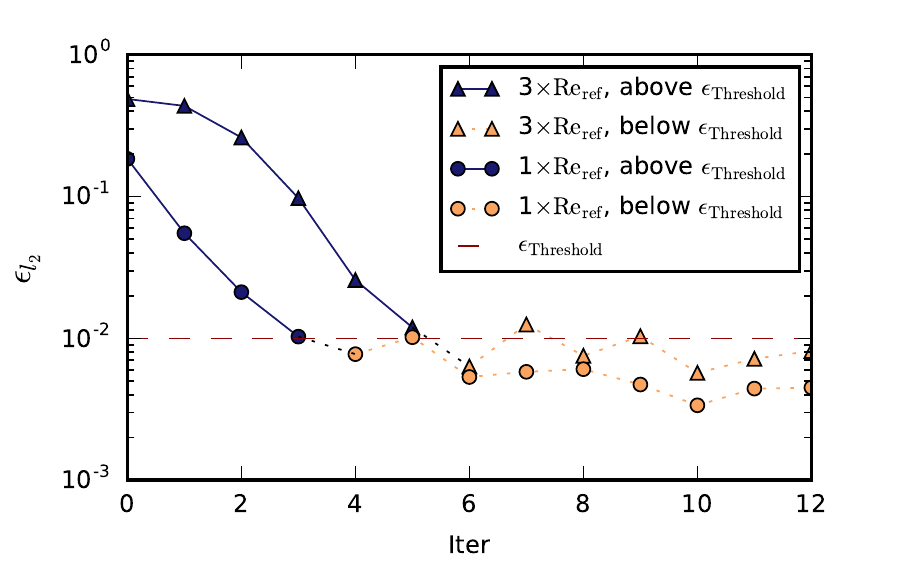} \label{l2error}} &
	\subfloat[]{\includegraphics[width=0.5\textwidth]{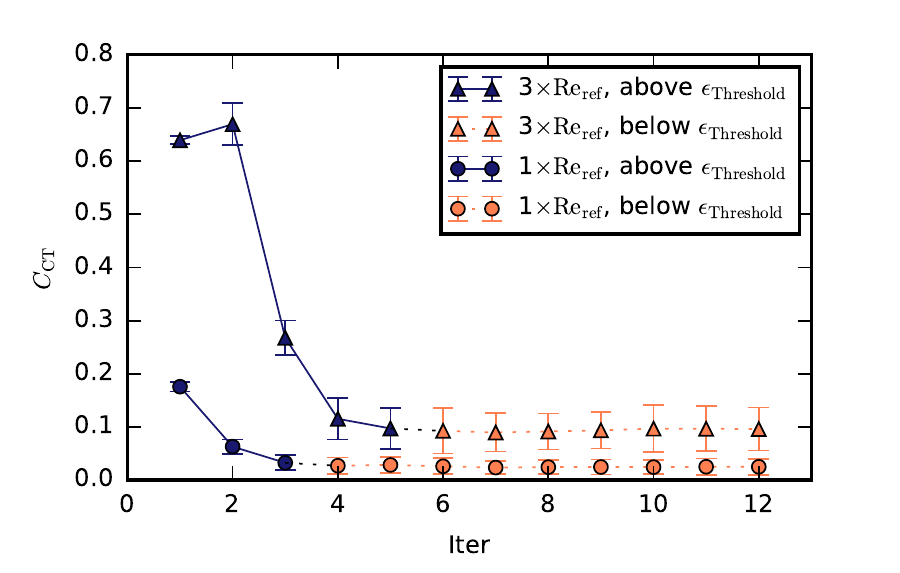} \label{Conver}}
	\end{tabular}
	\caption{(a) Rate of convergence of a two-dimensional cavity flow with \AlgAbb. The normalized error $\epsilon_{l_2}$ at $T_\text{final}$ is plotted for two different estimated Reynolds number.  (b) The corresponding termination condition $C_\text{TC}$ is calculated according to Eq.~\eqref{eqn:term_condition}. Solid navy lines represent the iterations before the error threshold, and dotted yellow lines after the error threshold.}
	\label{fig:Cavity_Convergence}
\end{figure}

A fluid region with a dimension of $30 \times 30 \times 40$ DPD units is surrounded by a solid wall consisted of fixed particles. At a particle density of $8$, we set the DPD parameter $\alpha=9.375$ and $\gamma=13.5$ to counter the effect of compressibility due to the fast moving wall at a velocity of $1.833$. This set of parameters results in a reference Reynolds number $\text{Re}_\text{ref}=200$. To create the no-slip cavity lid and wall, we used an effective dissipative coefficient for liquid–solid interactions~\cite{li2018dissipative}. For the cavity flow, we used the same projection method as in the Poiseuille flow example, with the consideration that the velocity profile consists of two components. As a result, both $x$ and $y$ components are recorded for each grid node. Because the grid in the coarse propagator spans in two dimensions, the mapping procedure for this example is slightly modified from that of plane-flow. The velocity of a particle is the linear interpolation of two nearest grid values. We used the same filtering techniques as in the prior examples.

Because an analytical solution does not exist for cavity flow, a reference solution is constructed by averaging the velocity fields from fifty independent serial DPD simulations. For the purpose of analysis, instead of particle velocity fields we inspect the equivalent stream function. The stream function can be easily obtained by solving the Poisson equation $-\nabla^2 \psi = ( \nabla \times \bm{u}) \cdot \hat{z}$. We then compute the normalized $l_2$ error $\epsilon_{l_2}$ between simulated and reference stream functions.

To test \AlgAbb's flexibility, we falsely assume that the fluid is much less viscous in the macroscopic model with a Reynolds number $\text{Re}_\text{est}=600$ which is $3\times$ the reference Reynolds number $\text{Re}_\text{ref}$. It is important to note that the falsely assumed viscosity would lead to an incorrect vortex location and, consequently, vastly different velocity field. As a result, the vortex from our \AlgAbb~simulation does not align with the reference vortex in early iterations shown in Fig~\ref{fig:cavity_diff}. The mis-alignment is then iteratively corrected. We again calculate $\epsilon_{l_2}$ at $T_\text{final}$ and plot it against iterations in Fig.~\ref{fig:Cavity_Convergence}\protect\subref{l2error}. The rate of convergence is no longer exponential as it was for the Poiseuille plane-flow. This is due to the notion that in two-dimensional space the $l_2$ error might not decay in a shortest path. For this particular case, it took 5 iterations to reach the threshold error $\epsilon_{\text{Threshold}}=1\%$ as shown in Fig.~\ref{fig:Cavity_Convergence}\protect\subref{l2error}. We also repeated the experiment with an estimated Reynolds number $\text{Re}_\text{est}$ that is equal to the reference $\text{Re}_\text{ref}$. As expected, it required fewer iterations to converge. The termination criterion was also computed and plotted in Fig.~\ref{fig:Cavity_Convergence}\protect\subref{Conver}. It exhibits the same shortcoming as in the Poiseuille flow - no clear separation between iterations above and below $\epsilon_{\text{Threshold}}$. Further discussion on this issue is presented in Section~\ref{sec:SumDis}.

\section{Summary \& Discussion}\label{sec:SumDis}
We have developed a supervised parallel-in-time algorithm for stochastic dynamics (\AlgAbb). As an extension to the Parareal algorithm, \AlgAbb~uses heterogeneous solvers, e.g., a Navier-Stokes solver as a predictor and the DPD method as a corrector, to accelerate stochastic particle simulations by decomposing the temporal domain. The predictor, also known as the coarse propagator, solves the macroscopic model, which is generally in the form of partial differential equations, in serial. Because the microscopic model converges to continuum macroscopic models in the scale limit, the time evolution of the macroscopic system then serves as an initial condition for all time subdomains. The corrector, also known as the fine propagator, recovers the micro-dynamics with expensive stochastic simulations in parallel. While here we demonstrated \AlgAbb~for hydrodynamics, the same framework can be applied to other problems, where long-time integration is hampered by the small characteristic time scale of the micro-dynamics.

To summarize, we first presented the \AlgAbb~algorithm and analyzed its theoretical speedup. We then discussed the importance of filtering and demonstrated its effect with an example problem. The accuracy and convergence of \AlgAbb~was subsequently tested.  We showed that the transient and final solutions match the analytical/reference solutions even when our estimations to the system’s mean-field behavior is far from the true behavior. More importantly, the \AlgAbb~algorithm is able to preserve the stochastic fluctuations of the microscopic model. Lastly, we demonstrated that \AlgAbb~provides better parallel efficiency and thus better scalability than the conventional domain decomposition method for long-time simulations.

There are three comments and suggestions we would like to make regarding \AlgAbb. First, although the solution converges regardless the value of estimated viscosity, an overly rough estimation requires a large number of iterations. As for the Poiseuille flow example, a rough estimate at~$10\times$ true viscosity requires~$20$ iterations, whereas only $5$ iterations are needed for a better estimate at~$2\times$ the true viscosity. In light of this correlation, we strongly encourage using an informed estimate of the macroscopic parameters in order to maximize the efficiency. Second, if the data from one microscopic state is insufficient to determine a mean-field quantity, the projection operation can be performed with data from multiple states with temporal-proximity. With the additional data, the mean-field quantity can be determined with higher precision. Lastly, we would like to discuss the issue of the termination condition. Given that \AlgAbb~is a general parallel-in-time algorithm, the termination condition presented in Section~\ref{sec:pasd} should be application-dependent. The stochastic noise is embedded in the refined solution. $C_{\text{Tolerance}}$ thus should be adjusted according to the magnitude of the noise. In the examples, we chose the normalized $l_2$ error as the quantity of interest $C_{\text{TC}}$. The termination criterion is satisfied when $C_{\text{TC}}$ falls below a threshold (see Eq.~\eqref{eqn:term_condition}). For the Poiseuille flow, we computed $C_{\text{TC}}$ for all iterations as shown in Fig.~\ref{fig:Poise}\protect\subref{fig:convthreshold}. When the parameter estimation is accurate as in the case of $2\times$ reference viscosity, $C_{\text{TC}}$ fluctuates mildly as indicated by the range of confidence interval and drops sharply to a level that satisfies the termination condition. However, when estimation is rough, the confidence interval is relatively large, which might cause the simulation to stop prematurely. In this case, we recommend using noise-filtered solution in the error calculation. Another option is to define a different quantity of interest as the termination criterion with a redefined termination condition. Finally, it is important to mention that the use of filtering helps to damp the high-frequency noises in the mean-field quantities, which only acts as a rough predictor for the fine stochastic solver. Regardless of the choice of filtering algorithms, the final solution obtained with \AlgAbb~contains all the statistics of the stochastic fluctuations and their correlations, consistent with unsupervised (serial in time) stochastic dynamics.

In future work, we would like to expand the applications of \AlgAbb~in two directions. First, we plan to simulate multi-physics dynamics along with hydrodynamics. For example, we can investigate stochastic non-isothermal systems at the microscale by incorporating the temperature equation in the macro-model. The temperature equation and the Navier-Stokes can be solved independently. For the micro-model, we can employ the energy dissipative particle dynamics, a flavor of DPD that tracks the energy of the system explicitly~\cite{2014ZLi_JCP}. Second, we will consider complex fluids that exhibit non-Newtonian behavior while employing a simple fluid model in the macro-scale. For example, we can model blood flow in the microscale, in which the non-Newtonian behavior can be captured with explicit representations of red blood cells and white blood cells~\cite{fedosov2010multiscale}. These are currently very expensive simulations and the aim is to test whether \AlgAbb~can speed up such simulations by orders of magnitude while maintaining the fidelity of microscale simulations.

\section*{Acknowledgements}
This work was supported by the DOE PhILMs project (No.~DE-SC0019453) and the U.S. Army Research Laboratory under Cooperative Agreement No.\ W911NF-12-2-0023. An award of computer time was provided by the ASCR Leadership Computing Challenge (ALCC) program. This research used resources of the Argonne Leadership Computing Facility, which is a DOE Office of Science User Facility supported under Contract DE-AC02-06CH11357. This research also used resources of the Oak Ridge Leadership Computing Facility, which is a DOE Office of Science User Facility supported under Contract DE-AC05-00OR22725. The authors would like to thank Dr. Fangying Song for providing technical support on data analysis and Prof. Yvon Maday for insightful discussions.

\bibliographystyle{elsarticle-num}
\biboptions{sort&compress}

\end{document}